\documentclass[twocolumn]{aastex631}

\usepackage{amsmath}

\begin{document}

\title{Equalizing the Pixel Response of the Imaging Photoelectric Polarimeter On-Board the IXPE Mission}

\correspondingauthor{John Rankin}
\email{john.rankin@inaf.it}

\author[0000-0002-9774-0560]{John Rankin}
\affiliation{INAF Istituto di Astrofisica e Planetologia Spaziali, Via del Fosso del Cavaliere 100, 00133 Roma, Italy}
\affiliation{Università di Roma Sapienza, Dipartimento di Fisica, Piazzale Aldo Moro 2, 00185 Roma, Italy}
\affiliation{Università di Roma Tor Vergata, Dipartimento di Fisica, Via della Ricerca Scientifica, 1, 00133 Roma, Italy}

\author[0000-0003-3331-3794]{Fabio Muleri}
\affiliation{INAF Istituto di Astrofisica e Planetologia Spaziali, Via del Fosso del Cavaliere 100, 00133 Roma, Italy}

\author[0000-0003-0331-3259]{Alessandro Di Marco}
\affiliation{INAF Istituto di Astrofisica e Planetologia Spaziali, Via del Fosso del Cavaliere 100, 00133 Roma, Italy}

\author[0000-0003-1533-0283]{Sergio Fabiani}
\affiliation{INAF Istituto di Astrofisica e Planetologia Spaziali, Via del Fosso del Cavaliere 100, 00133 Roma, Italy}

\author[0000-0001-8916-4156]{Fabio La Monaca}
\affiliation{INAF Istituto di Astrofisica e Planetologia Spaziali, Via del Fosso del Cavaliere 100, 00133 Roma, Italy}

\author[0000-0002-7781-4104]{Paolo Soffitta}
\affiliation{INAF Istituto di Astrofisica e Planetologia Spaziali, Via del Fosso del Cavaliere 100, 00133 Roma, Italy}

\author[0000-0002-4576-9337]{Matteo Bachetti}
\affiliation{INAF/Osservatorio Astronomico di Cagliari, Via della Scienza 5, I-09047 Selargius (CA), Italy}

\author[0000-0002-9785-7726]{Luca Baldini}
\affiliation{Università di Pisa, Dipartimento di Fisica Enrico Fermi, Largo B. Pontecorvo 3, I-56127 Pisa, Italy}
\affiliation{Istituto Nazionale di Fisica Nucleare, Sezione di Pisa, Largo B. Pontecorvo 3, I-56127 Pisa, Italy}

\author[0000-0003-4925-8523]{Enrico Costa}
\affiliation{INAF Istituto di Astrofisica e Planetologia Spaziali, Via del Fosso del Cavaliere 100, 00133 Roma, Italy}

\author[0000-0002-7574-1298]{Niccolò Di Lalla}
\affiliation{W.W. Hansen Experimental Physics Laboratory, Kavli Institute for Particle Astrophysics and Cosmology, Department of Physics and SLAC National Accelerator Laboratory, Stanford University, Stanford, CA 94305, USA}

\author[0000-0002-0998-4953]{Alberto Manfreda}
\affiliation{Istituto Nazionale di Fisica Nucleare, Sezione di Pisa, Largo B. Pontecorvo 3, I-56127 Pisa, Italy}

\author[0000-0002-1868-8056]{Stephen L. O'Dell}
\affiliation{NASA Marshall Space Flight Center, Huntsville, AL 35812, USA}

\author[0000-0003-3613-4409]{Matteo Perri}
\affiliation{INAF/Osservatorio Astronomico di Roma, Via Frascati 33, I-00040, Monte Porzio Catone (RM)}

\author[0000-0002-2734-7835]{Simonetta Puccetti}
\affiliation{Agenzia Spaziale Italiana, Via del Politecnico snc, I-00133 Roma, Italy}

\author{Brian D. Ramsey}
\affiliation{NASA Marshall Space Flight Center, Huntsville, AL 35812, USA}

\author[0000-0001-5676-6214]{Carmelo Sgrò}
\affiliation{Istituto Nazionale di Fisica Nucleare, Sezione di Pisa, Largo B. Pontecorvo 3, I-56127 Pisa, Italy}

\author[0000-0002-9443-6774]{Allyn F. Tennant}
\affiliation{NASA Marshall Space Flight Center, Huntsville, AL 35812, USA}

\author[0000-0001-9194-9534]{Antonino Tobia}
\affiliation{INAF Istituto di Astrofisica e Planetologia Spaziali, Via del Fosso del Cavaliere 100, 00133 Roma, Italy}

\author[0000-0002-3180-6002]{Alessio Trois}
\affiliation{INAF/Osservatorio Astronomico di Cagliari, Via della Scienza 5, I-09047 Selargius (CA), Italy}

\author[0000-0002-5270-4240]{Martin C. Weisskopf}
\affiliation{NASA Marshall Space Flight Center, Huntsville, AL 35812, USA}

\author[0000-0002-0105-5826]{Fei Xie}
\affiliation{Guangxi Key Laboratory for Relativistic Astrophysics, School of Physical Science and Technology, Guangxi University, Nanning 530004, China}
\affiliation{INAF Istituto di Astrofisica e Planetologia Spaziali, Via del Fosso del Cavaliere 100, 00133 Roma, Italy}

\begin{abstract}

The Gas Pixel Detector is a gas detector, sensitive to the polarization
of X-rays, currently flying on-board IXPE --- the first observatory
dedicated to X-ray polarimetry. It detects X-rays and their polarization by imaging the ionization
tracks generated by photoelectrons absorbed in the sensitive volume, and then reconstructing the initial direction of the photoelectrons. The primary ionization charge is multiplied and ultimately collected on a finely-pixellated ASIC specifically developed for X-ray polarimetry. The signal of individual pixels is processed independently and gain variations can be substantial, of the order of 20\%. Such variations need to be equalized to correctly
reconstruct the track shape, and therefore its polarization direction. The method to do such equalization is presented here and is based on the comparison between the mean charge of a pixel with respect to the other pixels for equivalent
events. The method is shown to finely equalize the response of the detectors on board IXPE, allowing a better track reconstruction and energy resolution, and can in principle be applied to any imaging detector based on tracks.

\end{abstract}


\section{Introduction}

The Imaging X-ray Polarimetry Explorer (IXPE) mission \citep{2021arXiv211201269W,2021AJ....162..208S}
is the first observatory dedicated to X-ray polarimetry. It carries
on board three X-ray optics modules and three X-ray polarization sensitive
detectors \citep{2021APh...13302628B} based on the Gas Pixel Detector
(GPD, \citet{2001Natur.411..662C,2006NIMPA.566..552B,2007NIMPA.579..853B}).
This detector has previously also flown on the PolarLight mission
\citep{2019ExA....47..225F} and will also fly on future missions
such as eXTP \citep{2019SCPMA..6229502Z}.

\begin{figure}
\begin{centering}
\includegraphics[width=0.8\linewidth]{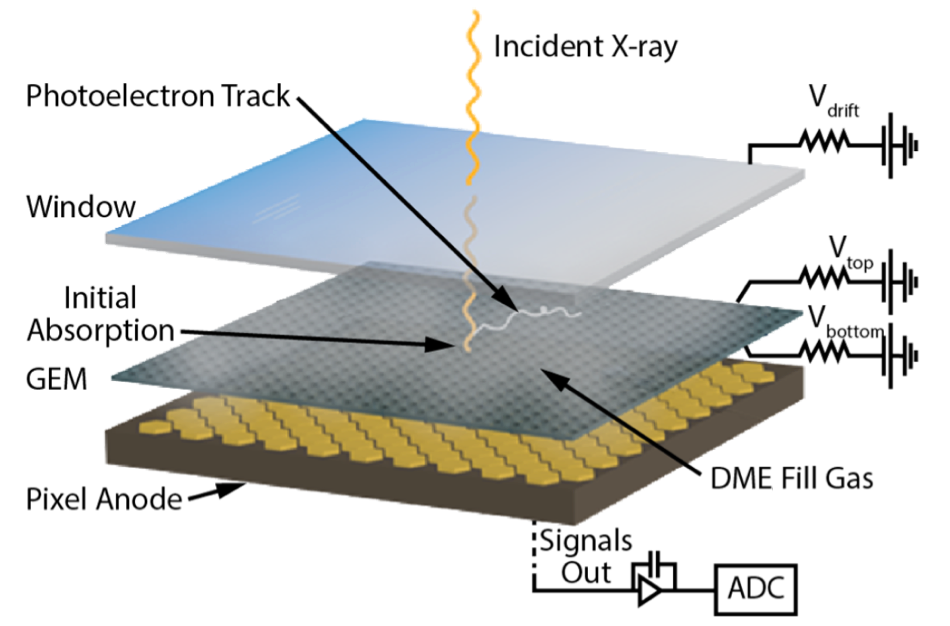}
\par\end{centering}
\caption{Schematic of the GPD. See the text for details. Image credit: \citet{2016SPIE.9905E..17W}
\label{fig:Schematic-GPD}}
\end{figure}

The GPD uses the photoelectric effect to measure the polarization of absorbed X-rays.
A schematic of it is shown in figure
\ref{fig:Schematic-GPD}: an incident X-ray enters through the beryllium
window and its absorption causes the formation of a photoionization
track in the dimethyl ether (DME) gas inside the cell. In the detector there is a drift
field, which causes the charge to move towards the Gas Electron Multiplier (GEM, \citet{2009NIMPA.608..390T}),
were it is multiplied. The charge is then collected by a custom ASIC,
specifically developed for X-ray polarimetry \citep{2006NIMPA.566..552B},
producing the image of the photoelectric track projected on the plane of the ASIC.

The azimuthal distribution of the directions of these tracks peaks along the
direction of polarization, while the amplitude of the modulation of
this distribution is proportional to the polarization degree. Therefore,
to compute the polarization of incident radiation, we need to determine
the direction of these ionization tracks. We do this using a reconstruction
algorithm based on the charge distributions of the image. Under development
is also a track reconstruction technique based on neural networks
\citep{2021NIMPA.98664740P}.

\begin{figure}
\begin{centering}
\includegraphics[width=1\linewidth]{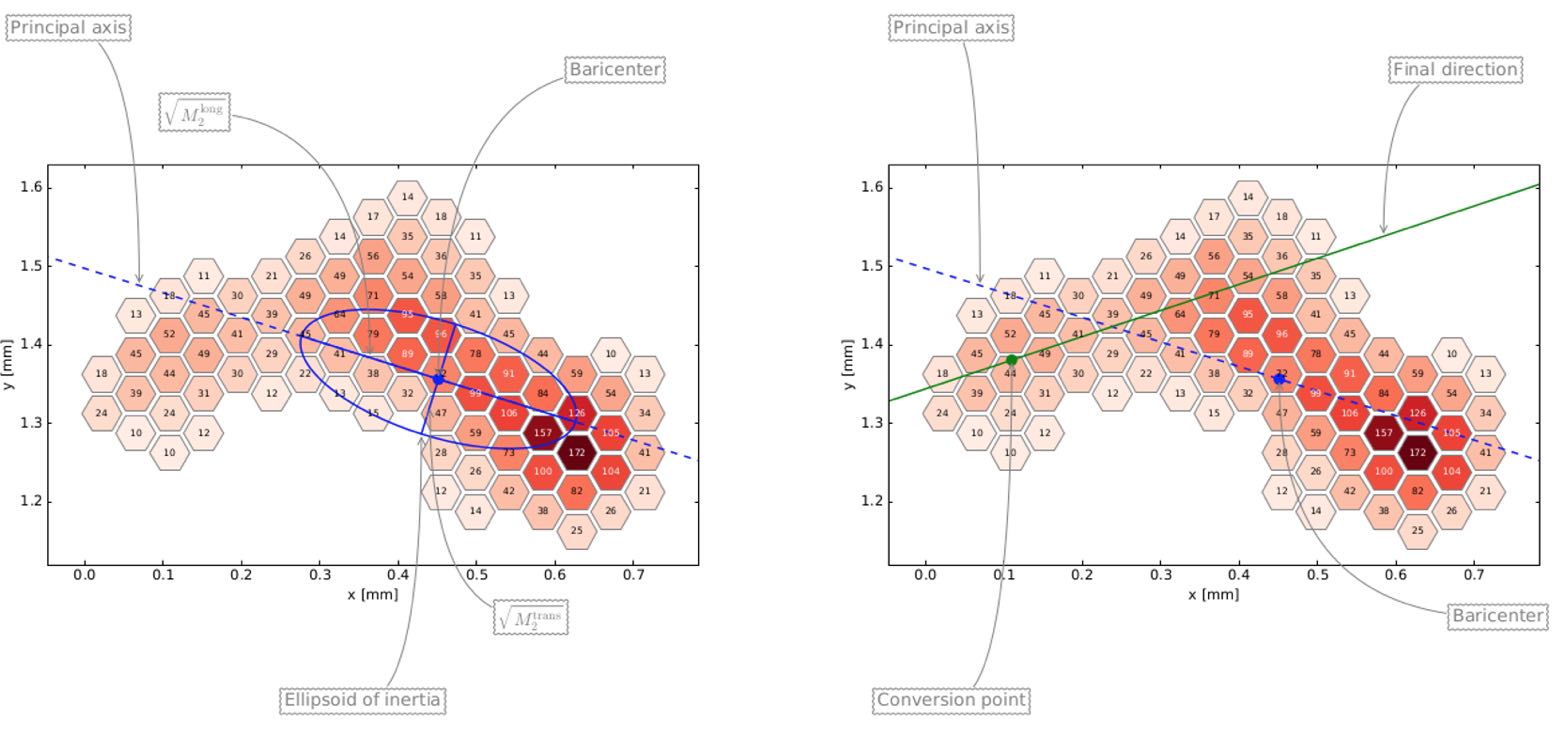}
\par\end{centering}
\caption{Example of a photoelectric track on the GPD. The part in darker color,
with the greatest charge density (the Bragg peak), is the one formed
last. The part with less charge is the one containing the real polarimetric
information. the reconstruction algorithm used to determine polarization
(\citet{2020JInst..15C4049M}) identifies the Bragg peak first (left
side of the figure) and then reconstructs the original direction (right
side of the figure).
\label{fig:Track}}
\end{figure}

Because the track direction contains all the polarization information,
its shape needs to be accurately imaged by the pixels of the ASIC,
requiring a sensitivity to measure signal of better than 1\%. This
is complex because Rutherford scattering causes the track
direction to vary during track formation. The example track image
in figure \ref{fig:Track} shows this: the
part with the greatest charge density (the Bragg peak) actually forms
last, while the part with the real polarimetric information forms
first and has got less charge. For this reason the reconstruction
algorithm that is used to determine polarization \citep{2003SPIE.4843..383B, 2014axp..book.....F, 2020JInst..15C4049M}
actually identifies the Bragg peak first and only at the second step
reconstructs the original direction. From the charge barycenter (only considering the first part of the track) the position is also computed, so that each event detected has a position associated to it even if the track is larger than the single pixels.

As is common in detectors made of pixels, the different pixels have
a different gain, which is a combination of the gain of the single ASIC pixel and of the gain of the GEM over that corresponding pixel. Separate calibration of all these items is not possible. Test capacitance for each pixel is available in the GPD ASIC, but their value has a large, unknown, variance, and so of little use for an electronic calibration. The gain of the GEM in different positions is not measured before the integration in the GPD, as the alignment between the GEM holes and the ASIC pixels is not guaranteed. As a consequence an equalization based on X-ray calibration measurements is necessary \citep{2022SPIE12181E..56R}.

In the following, we will associate pixel-to-pixel variations to variations in the pixel feedback capacitance, while large-scale structures will be related to variations in the GEM gain (see \citet{2021APh...13302628B}). 
Both of these variations cause disuniformities in the image which can
change the reconstructed direction of all the ionization tracks, potentially
introducing a systematic effect in the measured polarization. 

In this paper we describe the application of a method
that, through the use of of ground calibration measurements, performs
the equalization of the combined gain of each pixel. The method requires calibration measurements with
great statistics, which were acquired for IXPE ground calibration;
nevertheless, the method can be applied to other detectors producing
images of tracks.

The result of our method is the gain response of the individual pixels. For the GPD ASIC, this can be fit with a linear relation, and therefore the results of the calibration is a map of the gain and the offset for all pixels. 
Each event acquired by the GPD consists of the image of a track; each
pixel of this image is calibrated, based on its position, by 
multiplying by the gain of these maps, so correcting the pixel gain. 
The track reconstruction
is done only after the image has been equalized.

This paper is structured as follows. In section \ref{sec:track-readout}
we described the GPD, with particular regard to the track readout
mechanism. In section~\ref{sec:Ground-calibration-measurements} we present the calibration measurements used as the input of the equalization procedure. In section \ref{sec:eq-method} we described the pixel
by pixel equalization method, and we present its applications to energy
resolution and to the response to polarization in section \ref{sec:Applying-the-method}.
Finally, in section \ref{sec:Conclusion}, we draw our conclusions.

\section{The track readout of the Gas Pixel Detector\label{sec:track-readout}}

The GPD ASIC is made of 300$\times$352 hexagonal pixels.
All pixels are collected in basic 2$\times2$ units of pixels, called
miniclusters. The four pixels are logically OR-ed together and have their own output chain.

\begin{figure}
\begin{centering}
\includegraphics[width=0.8\linewidth]{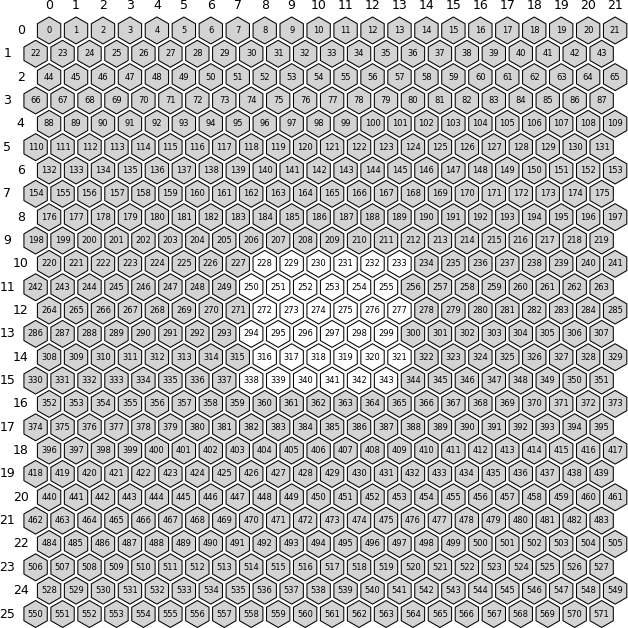}
\par\end{centering}
\caption{Region Of Interest (ROI) made of the pixels that have triggered
(white) plus the other pixels --- 8 (4 miniclusters) in the x direction and 10 (5 miniclusters) in
the y direction --- that are read (grey). 
\label{fig:roi_pixels}}
\end{figure}

The role of the minicluster is to issue a trigger when the collective
signal from the four pixels exceeds a certain threshold, which can
be adjusted for each minicluster. As the charge density of the photoelectron
track is higher at the Bragg peak, the miniclusters usually trigger
in the end part of the track.
For this reason, when an event occurs, a padding in each coordinate is introduced so to read the whole track --- including its initial part: the pixels read are these from
a region called \emph{Region Of Interest (ROI)}, which is composed
of the miniclusters that triggered, plus a padding of 8 pixels (4 miniclusters)
along the x direction and 10 pixels (5 miniclusters) along the y direction.
The reason for the discrepancy between the two axes is to compensate
the fact that each minicluster has not a square shape but, due to
the hexagonal shape of the pixels, the width is greater than the height.
An exemplary ROI is shown in figure \ref{fig:roi_pixels}; ROIs are generated by the ASIC starting from the largest rectangular region that contains the miniclusters that have triggered, with the padding added to this rectangle.
This property is used so that not all pixels of the detector are read,
but only these in the region around the track, rendering the readout
of a track much faster.

\begin{figure}
\begin{centering}
\includegraphics[width=0.8\linewidth]{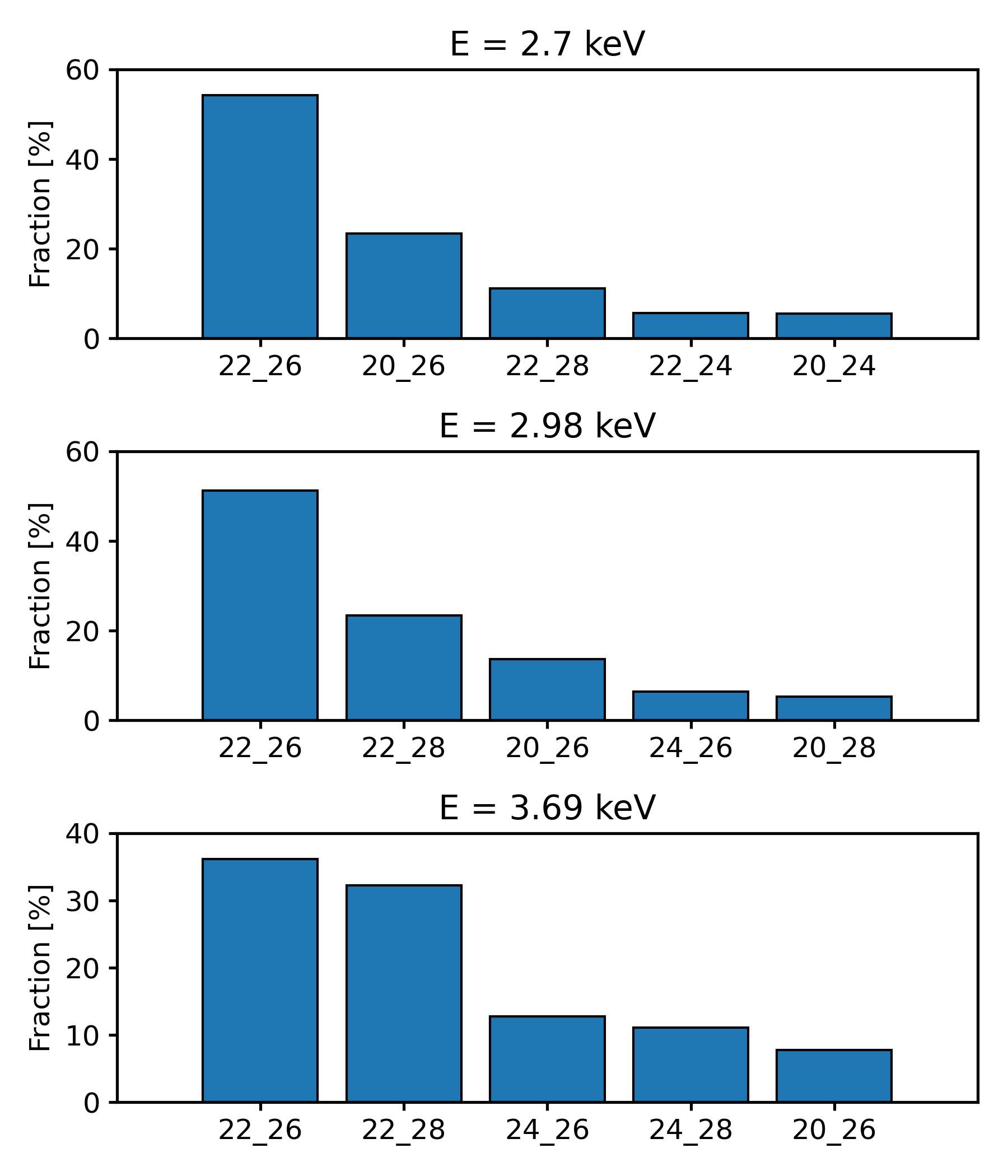}
\par\end{centering}
\caption{Distribution of different ROI sizes at different energies. The values on the x axis indicate the size of the ROI (its number of pixels) respectively over the x and y dimension. \label{fig:roi_size_hist}}
\end{figure}

\begin{figure}
\begin{centering}
\includegraphics[width=1\linewidth]{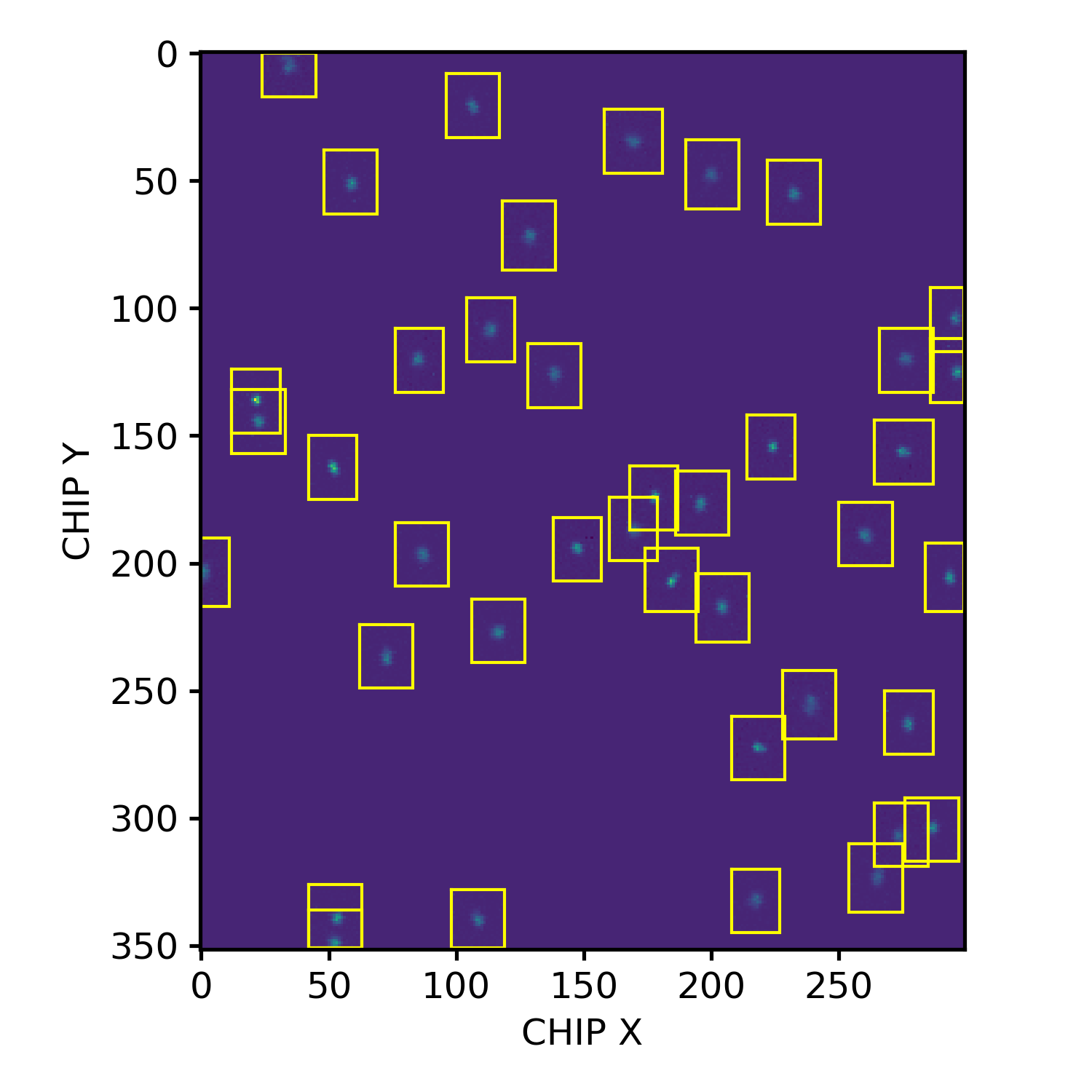}
\par\end{centering}
\caption{Example acquisition with many different tracks and corresponding ROIs
over the entire surface of the ASIC. \label{fig:rois_over_detector}}
\end{figure}

Because the number of miniclusters that trigger is variable, there
is a wide range of possible sizes for the ROIs. The smallest possible
is when only one minicluster is triggered, and in this case the ROI
size is $(1+4\cdot2)\times(1+5\cdot2)=9\times11\;\text{miniclusters}=18\times22\;\text{pixels}$.
For tracks produced by X-rays of higher energy, the number of miniclusters
that trigger, and therefore the ROI sizes, tend to be larger. ROIs
larger than a threshold of 4096 pixels are rejected; these are generated
by background events which are not of interest for X-ray polarimetry,
and would fruitlessly increase the detector deadtime. The distribution
of the most common ROI sizes at different energies is shown in figure
\ref{fig:roi_size_hist}. An example acquisition with many different
tracks and corresponding ROIs, over the entire surface of the ASIC,
is shown in figure \ref{fig:rois_over_detector}.


\newpage

\section{Ground calibration measurements\label{sec:Ground-calibration-measurements}}

The GPDs on board IXPE were extensively calibrated on ground before
launch \citep{2018SPIE10699E..5CM,ixpe_cal_sergio,MULERI2022102658} ---
with the calibration equipment developed at INAF-IAPS \citep{MULERI2022102658}.
Various types of measurements were performed; most of the time was
dedicated to acquiring monochromatic unpolarized and polarized measurements.
Each calibration measurement was repeated in two configurations: as
a flat field illuminating uniformly the entire surface of the detector, and as a
so-called deep flat field only in the central area (3.3~mm circular
radius) of the detector --- which was therefore calibrated with greater
statistics. The unpolarized measurements were acquired at 6 energies
(2.04, 2.29, 2.7, 2.98, 3.69, 5.89~keV), and the polarized ones at
7 energies (2.01, 2.29, 2.7, 2.98, 3.69, 4.51, 6.4~keV). Each measurement
was repeated two times at two orthogonal angles, so to decouple the
intrinsic response of the instrument to unpolarized radiation and measure
the source polarization (see \citet{2022AJ....163...39R}). More details
on the calibration products of this method is reported below in section
\ref{sec:Applying-the-method}.

Among the available measurements, we used for pixel equalization those
with the most statistics: the unpolarized flat
fields and deep flat fields at 2.7, 2.98 and 3.69~keV. The measurements
at the two orthogonal angles were summed to accumulate statistics:
as the polarization of the source is very low (Ratheesh et al. in preparation), the systematic differences
in the tracks were minor.

The calibrations were identical for the three flight detectors and
the additional spare unit. In the following, we will first introduce the method using the data
taken from the calibrations of IXPE detector unit number 2; after that we will compare how pixel equalization affects the three detectors.

\section{Pixel by Pixel equalization method\label{sec:eq-method}}

In Section \ref{sec:track-readout} we have seen that, from a certain input charge distribution, a number of miniclusters trigger, defining a ROI with a specific size and position. Each physical pixel in the ROI will collect a charge which depends on the gain at that physical position and on its position with respect to the input charge distribution. If instead we average all ROIs with the same size and position the charge collected by each pixel will depend only on its gain and on its relative position in the average ROI, and no more on its position with respect to the input charge, which varies wildly from a ROI to another. This method implicitly assumes that the triggering of the ROI is reasonably uniform; the good results of the discussed method confirm such an assumption.

\begin{figure}
\begin{centering}
\includegraphics[width=1\linewidth]{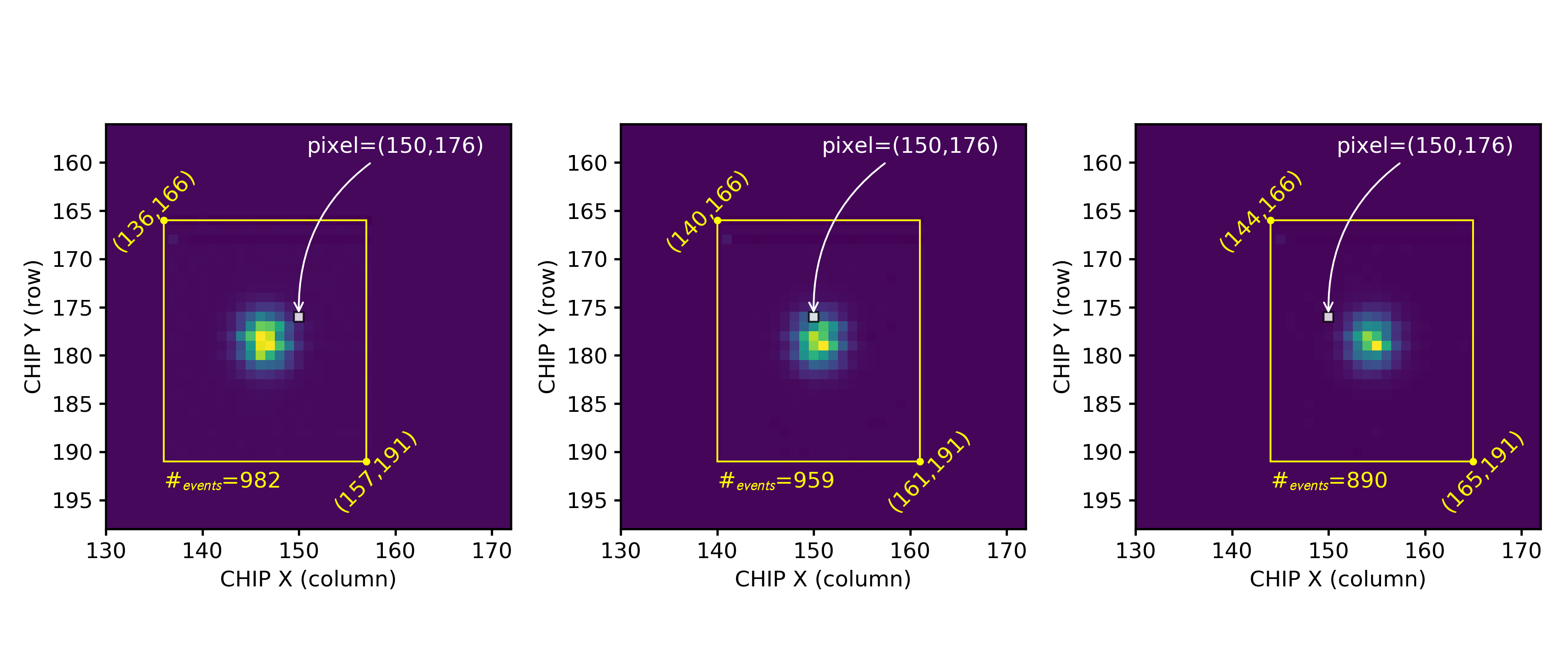}
\par\end{centering}
\caption{The charge a pixel collects in a certain ROI depends both on its gain
and on its relative position inside the ROI (yellow rectangle). In
all three cases the ROI has size 22$\times$26 (the most abundant
size), and is the average of a different number of events ($\#_{events}$)
of the same size and at the same position. \label{fig:pixel_different_rois}}
\end{figure}

Figure \ref{fig:pixel_different_rois} shows a few average ROIs with the same size (22$\times$26, which is the most abundant size) but at different physical positions. 
The same physical pixel can be outside the track (subplot on the right), at the border (subplot on the left) and inside the track (subplot at the center). 
Because the physical pixel is the same, the gain is the same, and the different charge collected by that pixel in the three ROIs depends on its position in the ROI.
This suggests that the gain of each pixel can be derived by systematically comparing the charge it collects with respect to the average of all other pixels in that same relative position in the average ROI.

\begin{figure}
\begin{centering}
\includegraphics[width=1\linewidth]{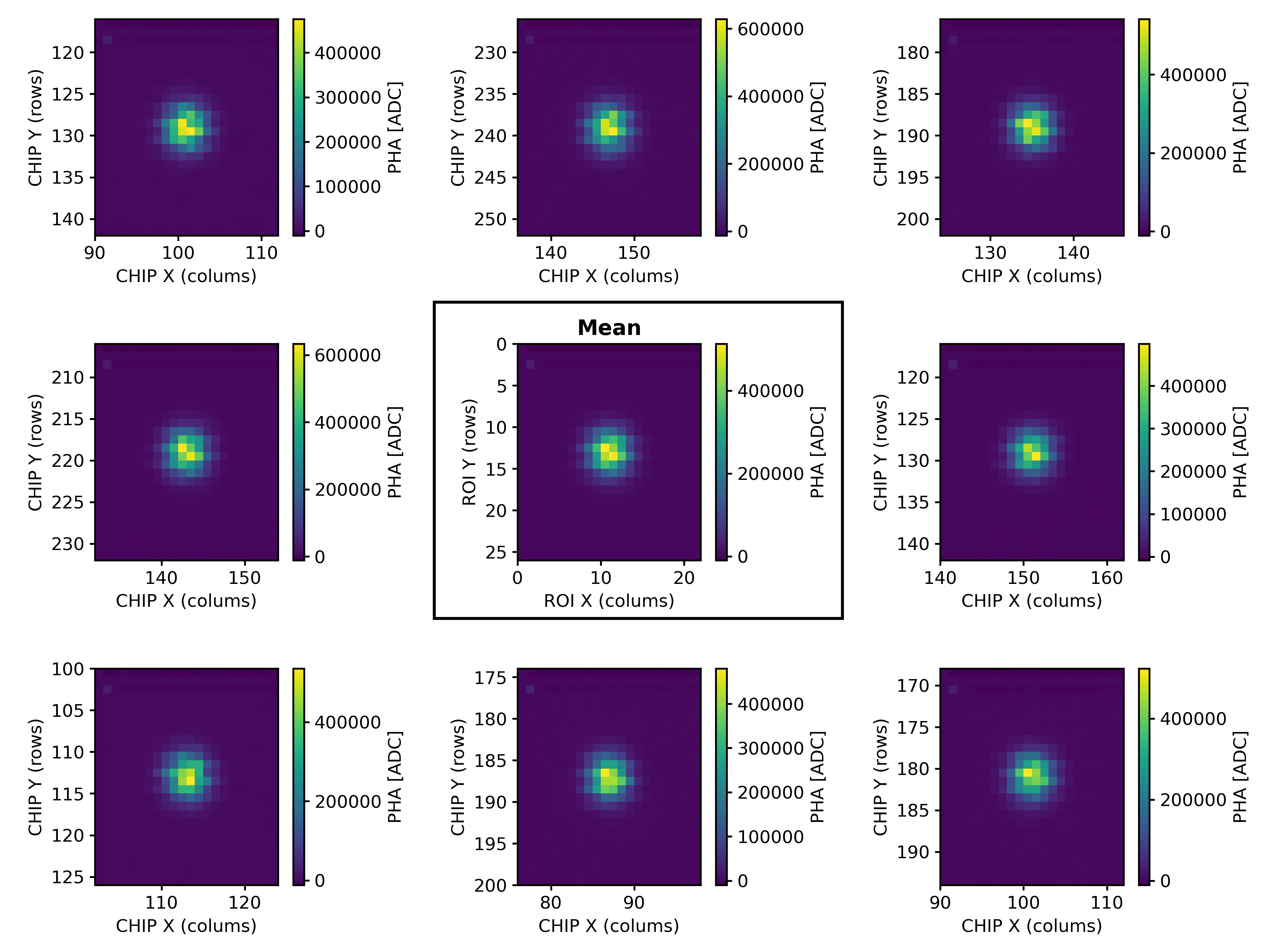}
\par\end{centering}
\caption{Example of different ROIs in different positions and their mean (central
plot). The tracks shown are from photons at 3.69~keV.
We do the averaging both to ROIs of the same shape and at the same position (to obtain \emph{average ROIs}), 
and also to ROIs of the same shape but at different positions (to obtain the \emph{reference ROI} for that shape).
\label{fig:rois_many_and_mean}}
\end{figure}

As a first step in the method we average ROIs of the same shape and at the same
position. This gives us an image for each ROI (defined based on its
start and end pixel positions in both the x and y axis) which we call \emph{average ROI}. We then average
ROIs of the same shape (with the same x and y size), but at different positions, to also obtain
the so-called \emph{reference ROIs}. Figure \ref{fig:rois_many_and_mean} shows an example
of how different ROIs can be averaged.

\begin{figure}
\begin{centering}
\includegraphics[width=1\linewidth]{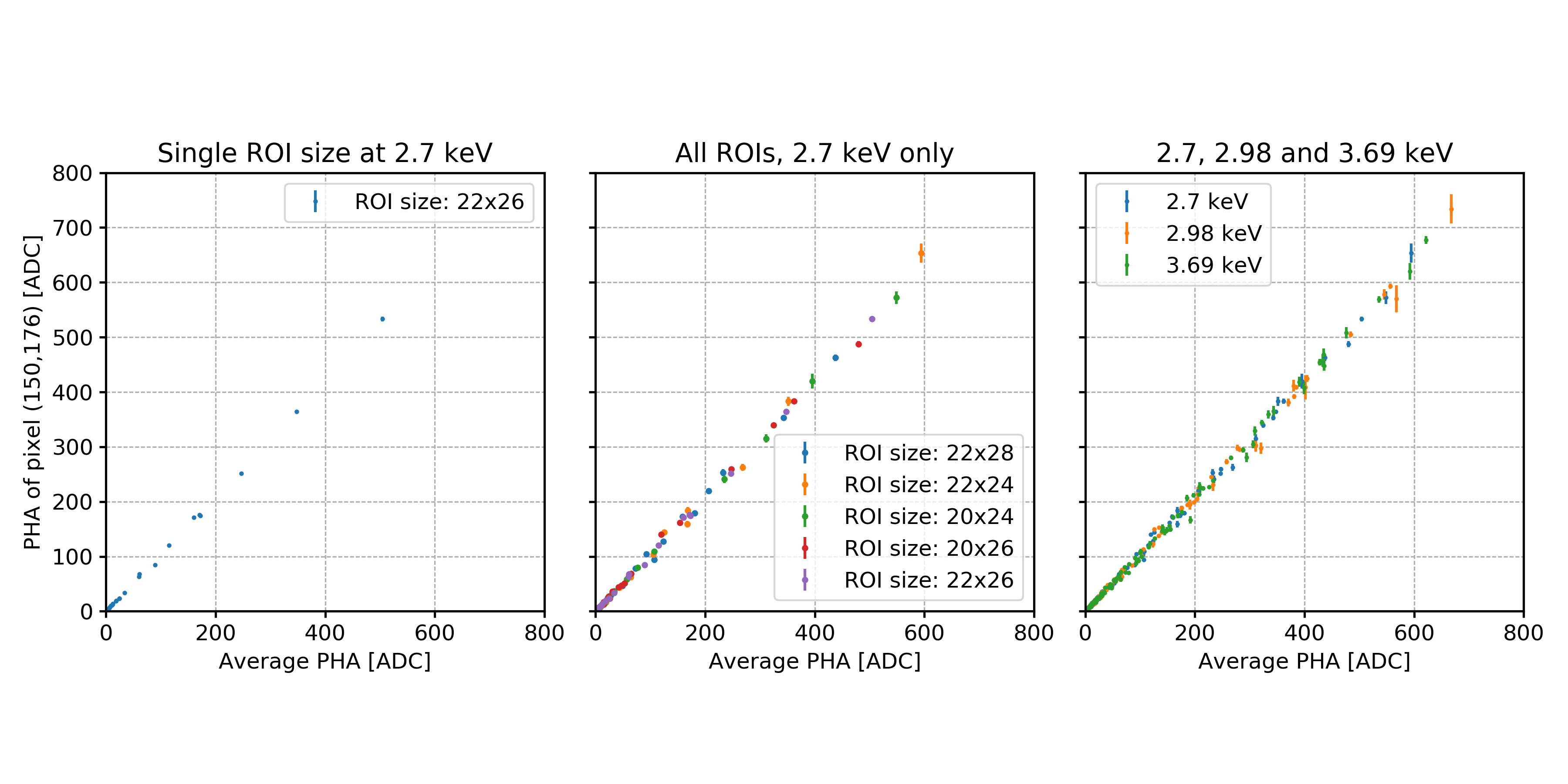}
\par\end{centering}
\caption{Single pixel response by comparing the charge in different ROIs at
the same absolute position (y axis) with the average charge at the
same relative position in the ROI (x axis). The left-most plot shows the pixel
response built with a single ROI size at a single energy, while the
subsequent plots show that the response built is unchanged if including
other ROI sizes and energies. \label{PHA_vs_PHA_3plots}}
\end{figure}

\begin{figure}
\begin{centering}
\includegraphics[width=1\linewidth]{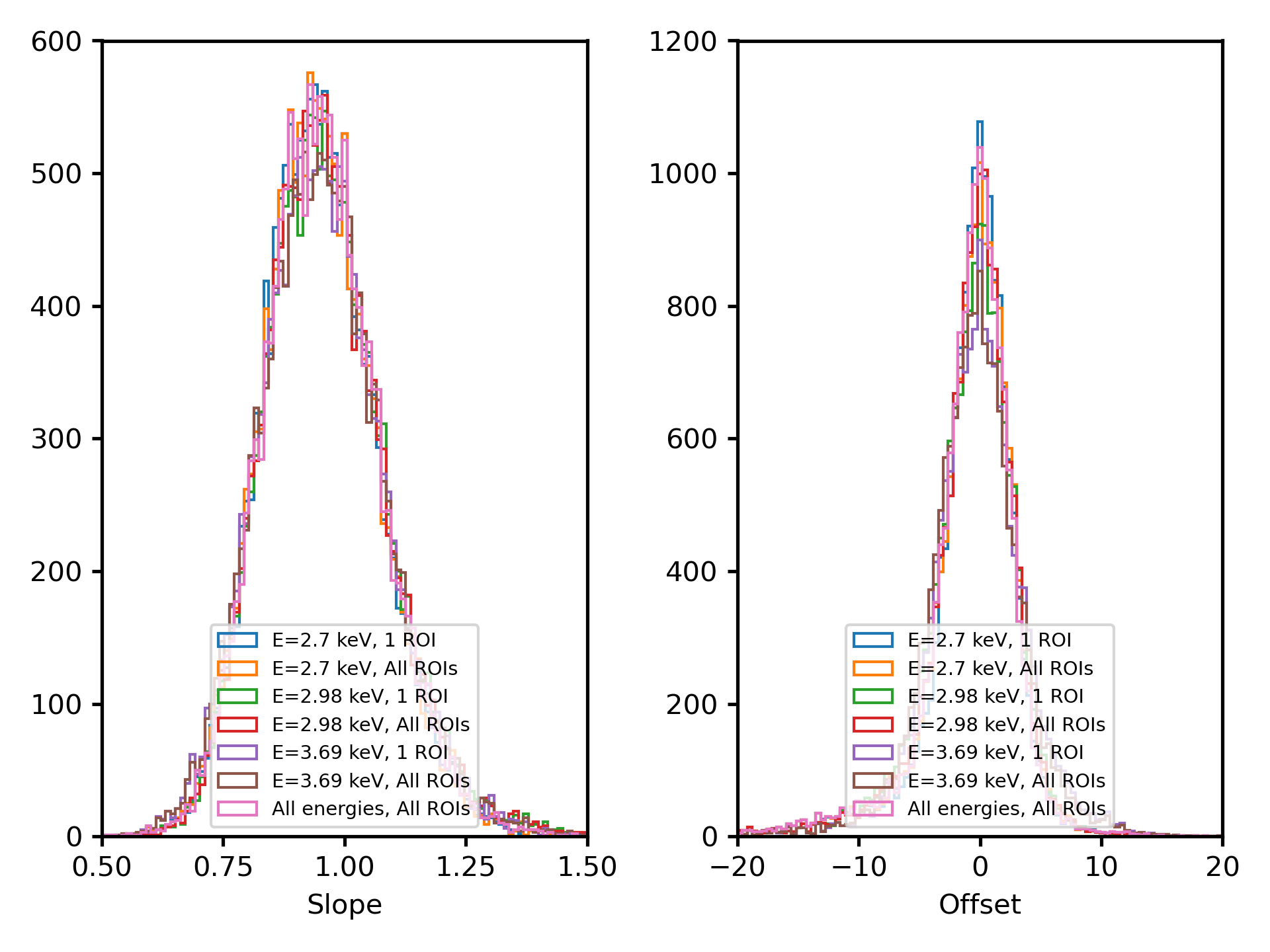}
\par\end{centering}
\caption{Distribution over all the pixels of the slope and offset of the linear fit of figure \ref{PHA_vs_PHA_3plots}
for different energies and ROIs and for their combination. When only
one ROI size is considered it's the one with the most counts.
\label{fit_roi_selection_hist_comparison}}
\end{figure}

The method is based on the comparison, for each pixel, of these
two kinds of averages. Figure \ref{PHA_vs_PHA_3plots} shows an example of ``pixel response''
for a specific pixel:
\begin{itemize}
    \item The values on the y axis is the charge collected by the pixel in different average ROIs. For any of them, the chosen pixel will occupy a different relative position $(i,j)$, e.g., the charge collected by the chosen pixels will be larger in the case of average ROIs which have it near their center.
    \item The value on the x axis is the charge of the reference ROI in the $(i,j)$ position. This represents the charge that, on average, the pixel would have to collect in the $(i,j)$ position.
\end{itemize}
The left-most plot shows the pixel response built with ROIs with a specific size (22$\times$26) and generated by (nearly) monochromatic photons at $\sim$2.7 keV. In the other panels of the figure, we add also the pixel response obtained by including events from other ROI sizes at the same energy, or ROIs obtained with measurements at other energies (see also figure \ref{fit_roi_selection_hist_comparison}). ROIs triggered by photons at different energies (or ROIs with the same size but generated from photons at different energies) have a different distribution of initial charge in the ROI, and then sample the pixel gain at different values. The very good correlation obtained with all these data-sets indicates that our working hypothesis --- that the charge distribution of average ROIs across the detector active area are different only because of the difference in pixel gain --- is adequate within the statistical uncertainties of our measurements.

Different ROI sizes occur
with different frequency (see figure \ref{fig:roi_size_hist}): to obtain good statistics,
for deep flat fields we use the 5 ROI sizes with the most occurrences,
while for flat fields we use the 2 with the most occurrences;
in both cases we also combine all 3 energies.

To obtain the response of all pixels we combine flat fields, covering the entire GPD sensitive area,
and deep flat fields (carried out in the central part only, see section \ref{sec:Ground-calibration-measurements})
in the following way. The central part (inside 3.3~mm radius) is obtained from deep flat fields which are calibrated with better sensitivity, while the gain of the outer pixels is obtained from flat fields. To avoid discontinuity among the two data-sets, the reference for each ROI size was extracted summing only the ROIs in the area common to the two measurements, that is, in the central 3.3~mm radius circle.

\begin{figure}
\begin{centering}
\includegraphics[width=1\linewidth]{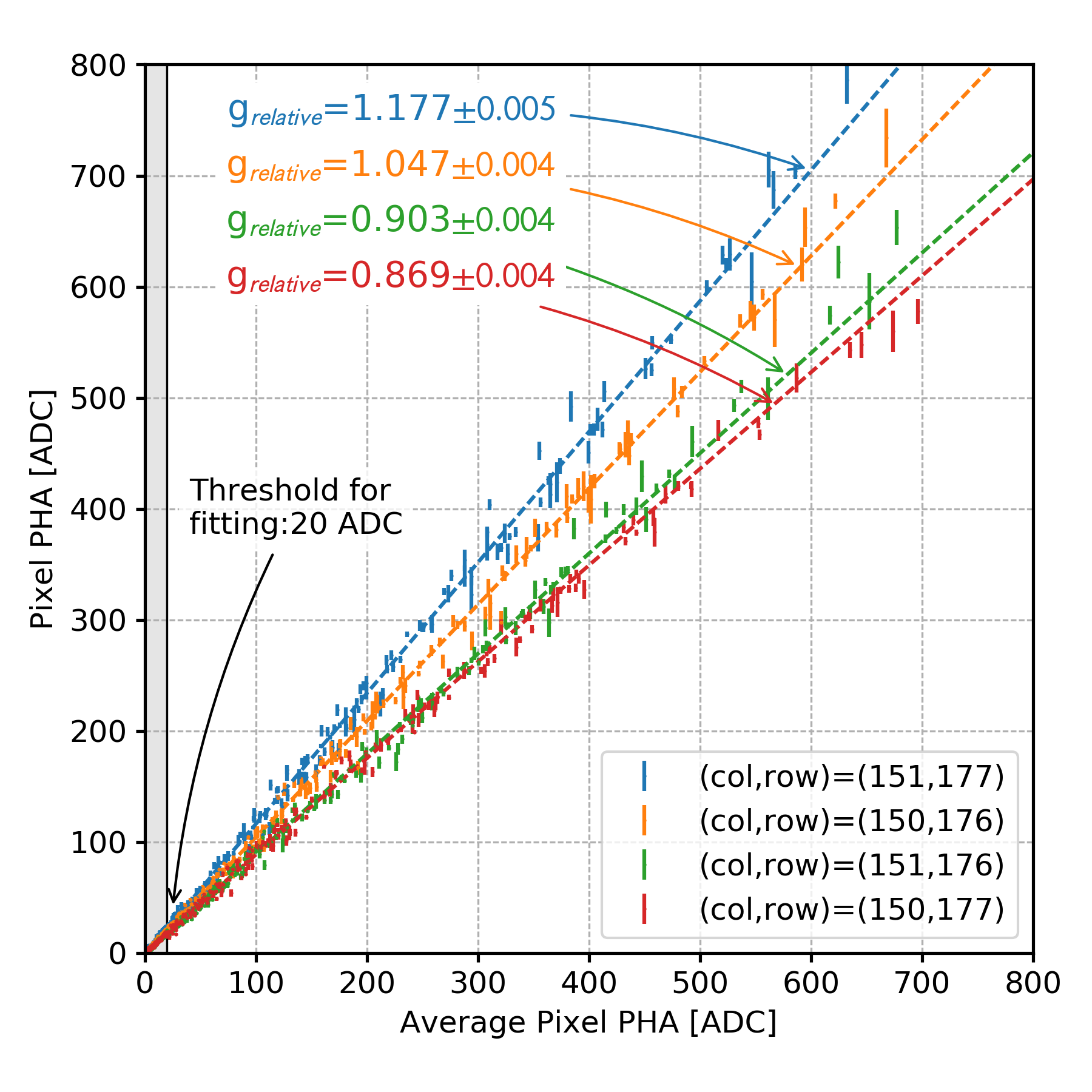}
\par\end{centering}
\caption{Same as figure \ref{PHA_vs_PHA_3plots} right, but for different pixels.
The variations are significant, even if the pixels are contiguous.
\label{PHA_vs_PHA}}
\end{figure}

We repeat the entire procedure for different pixels (figure \ref{PHA_vs_PHA}),
therefore obtaining the pixel response over all the detector; the
figure shows neighboring pixels: variations are significant. 

\begin{figure}
\begin{centering}
\includegraphics[width=1\linewidth]{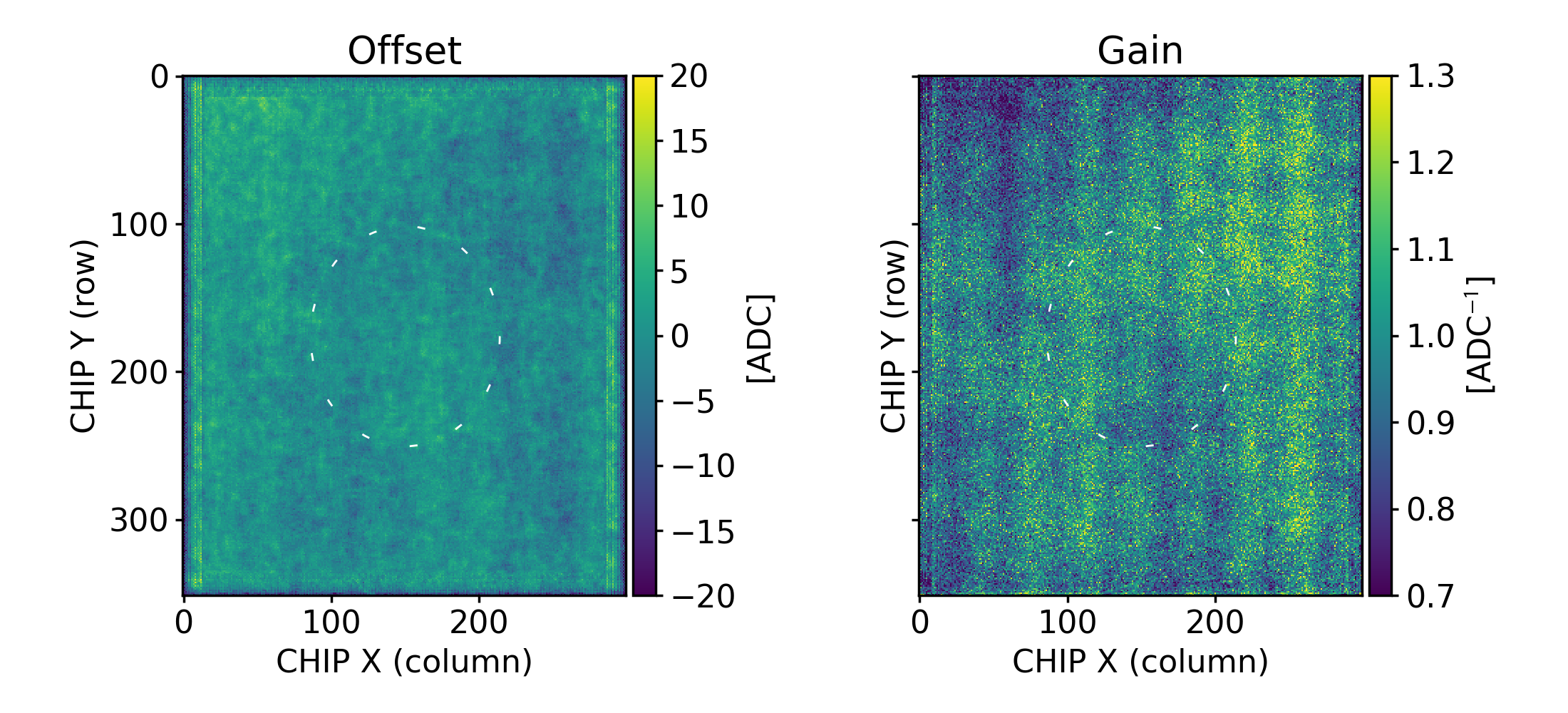}
\par\end{centering}
\caption{Maps for the offset (left) and gain (right) terms of the linear fit
of the quantities plotted in figure \ref{PHA_vs_PHA}, for the IXPE detector 2. For the central
area calibration measurements with greater statistics are available:
the boundary between the two regions is the white ellipsis. Some stripes
are visible, and are due to the average GEM gain (see \citet{2021APh...13302628B}).
\label{fig:Maps-linear}}
\end{figure}

We fit the relation of figure \ref{PHA_vs_PHA} with a linear function
to obtain the response for each pixel. Figure \ref{fig:Maps-linear}
shows the maps of the fitted coefficients. 
These are the maps saved in the calibration database used by IXPE's data reduction
pipeline. 

\section{Testing the method\label{sec:Applying-the-method}}

The pixel equalization is applied by multiplying the charge $\text{PHA}[i,\text{CHIPX}, \text{CHIPY}]$ of each pixel of each event by the gain term of the equalization maps (figure \ref{fig:Maps-linear}) 
\begin{multline*}
    \text{PHA\_EQ} [i, \text{CHIPX}, \text{CHIPY}] = \\
    \text{PHA}[i, \text{CHIPX}, \text{CHIPY}] \cdot \text{Gain}[\text{CHIPX}, \text{CHIPY}]
\end{multline*}
The offset term is negligible (a few bins of the analog-to-digital converter channels (ADC), compared to hundreds of ADCs
for the pixel charge), as expected because pixel pedestals are read-out and subtracted in real time after the event. Therefore, for simplicity, the IXPE data reduction pipeline
only applies the gain term.
In this section we compare how the spectral capabilities of the GPD,
and the response to polarization, change when applying the pixel equalization.

\subsection{Spectral capabilities}

The charge of each event measured by the GPD is the sum of the charges
of all the track pixels, and is proportional to the energy. 
The charge is initially expressed in arbitrary units or Pulse Height Amplitude (PHA), and is subsequently converted to the so-called Pulse Invariant (PI) energy channels, which go from 0 to 374 for energies respectively from 0~keV to 15~keV.
The energy resolution is computed as the ratio of the FWHM of the spectral line over its peak
energy. 

\begin{figure}
\begin{centering}
\includegraphics[width=1\linewidth]{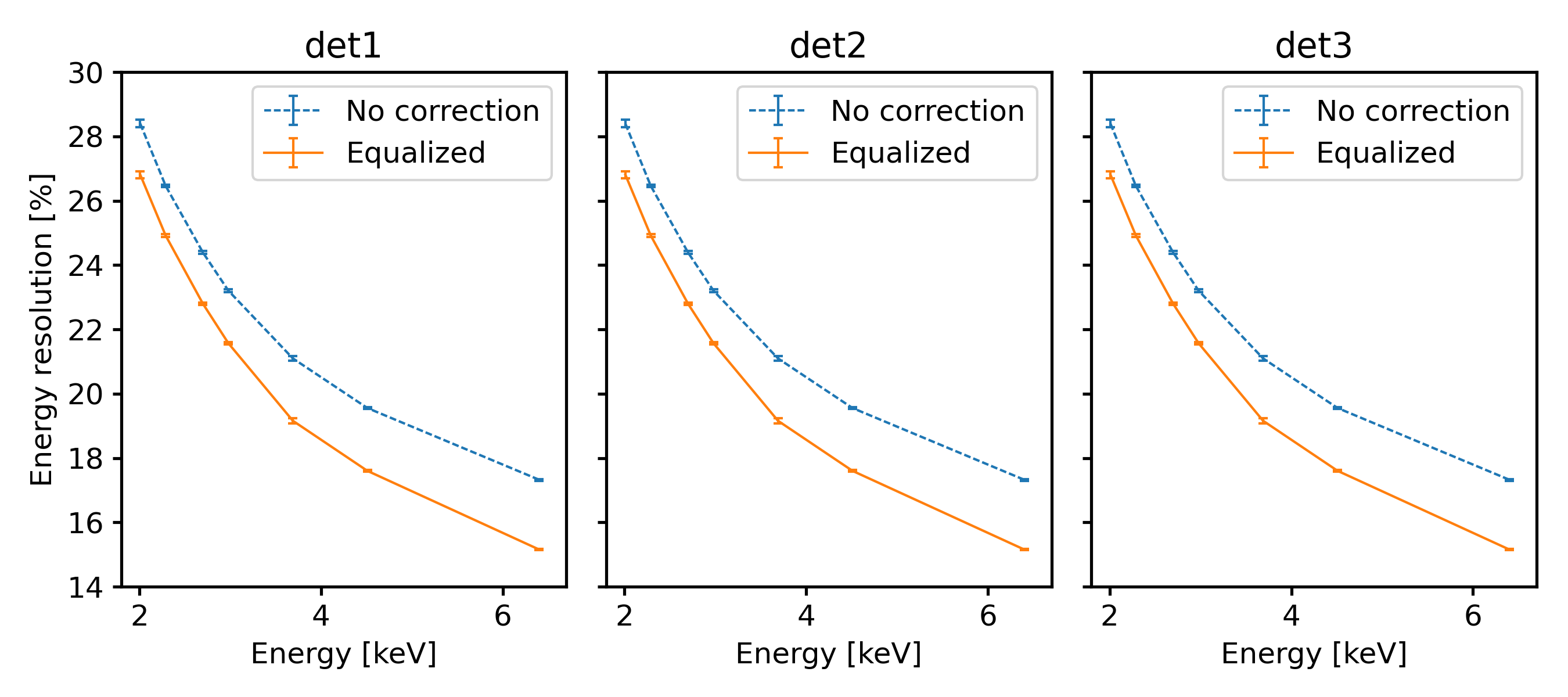}
\par\end{centering}
\caption{Energy resolution, as a function of energy, in a 1.5~mm radius spot
at the center of the three detectors on-board IXPE, computed without and with the application
of pixel equalization. \label{fig:Energy-resolution}}
\end{figure}

Figure \ref{fig:Energy-resolution} shows the energy resolution as a function of energy, using events in a 1.5~mm
radius spot at the center of the detector. This is the same radius at which, to reduce systematic effects, observations with IXPE are dithered \citep{2021arXiv211201269W}.
As expected, there is an improvement when applying the pixel equalization,
with the improvement increasing with energy, in a comparable way for the three detectors.

\begin{figure}
\begin{centering}
\includegraphics[width=1\linewidth]{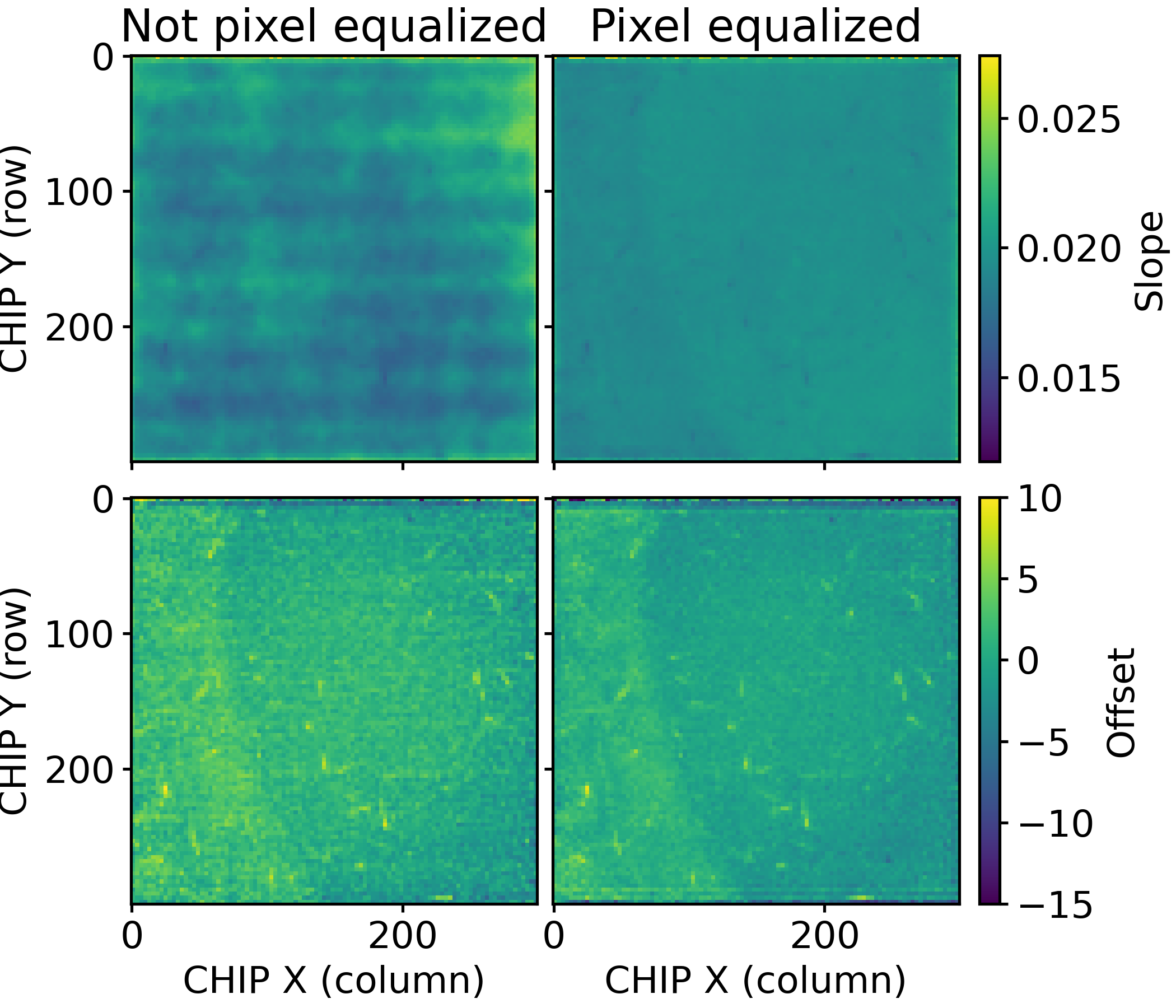}
\par\end{centering}
\caption{Slope and offset terms of the relation between Pulse Height Spectra
(PHA) and Pulse Invariant (PI) energy channels,
generated starting from non-equalized and equalized data, in $100\times100$ spatial bins, for the IXPE detector 2.
After equalization, large-scale variations are much less evident; remaining effects are eventually corrected with another calibration procedure. 
The tracks visible in the offset maps, instead, are due to alpha particles produced by residual radioactivity of the Beryllium window. In fact, the large energy deposit they release reduces temporarily the gain of the GEM in the region which is involved in the multiplication \citep{2021APh...13302628B}.
\label{fig:pkgain}}
\end{figure}

\begin{figure}
\begin{centering}
\includegraphics[width=1\linewidth]{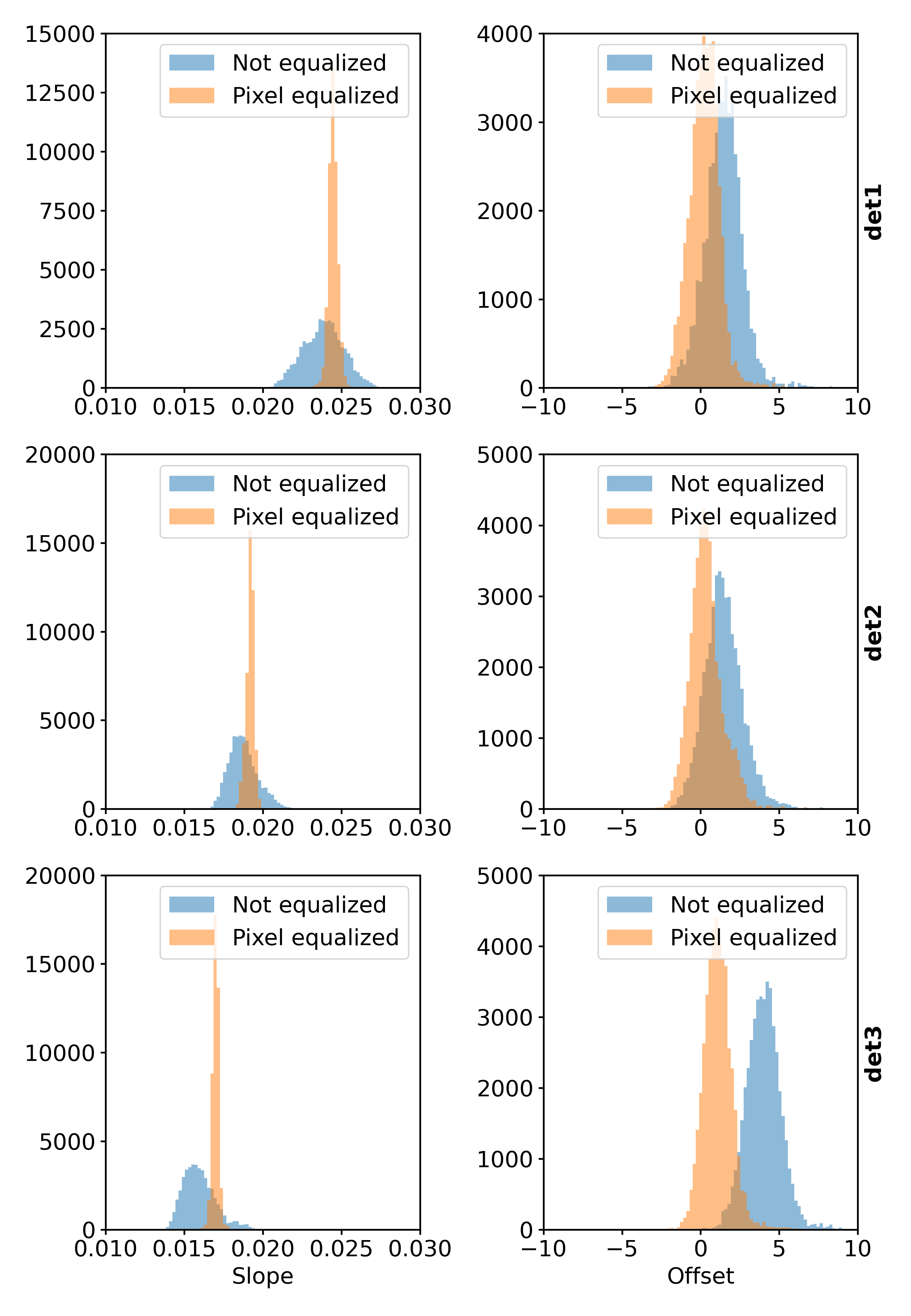}
\par\end{centering}
\caption{Same as figure \ref{fig:pkgain} but as histogram, and for all three detectors flying on-board IXPE. The results for the three detectors are comparable. \label{fig:pkgain-hist}}
\end{figure}

The relation between the measured energy (expressed in PHA) and the expected value (expressed in PI and, then, in keV) is derived during GPD calibration by using monochromatic sources at different and known energies. The relation between these two quantities can be assumed to be linear for the GPD, and we show in figures \ref{fig:pkgain} and \ref{fig:pkgain-hist} the map and the histograms of the derived slope and offset values in $100\times100$ independent spatial bins (defined from the barycenters of the tracks of each event) over the entire active area of the detector. The values are plotted both including or not the pixel equalization: its effect is to strongly reduce the variance of the slope values and most of the structures due to the GEM, see \citet{2009NIMPA.608..390T,2021APh...13302628B}. Remaining large-scale variations are eventually calibrated by normalizing the peak of sources of monochromatic photons (see \citet{2016SPIE.9905E..4GM}).

\subsection{Response to polarization}

Pixel equalization is expected to impact on the reconstruction of
single tracks, which are significantly distorted by variations of $\pm$20\%
in the pixel gain. 
In this section we analyze the impact of pixel equalization on the
polarization response, by comparing the modulation factor and spurious
modulation with and without the application of the pixel equalization.
The modulation factor is the response of the detector to 100\% polarized
radiation, and it is the factor over which the modulation must be
divided to obtain the polarization. Spurious modulation is the response
of the detector to unpolarized radiation, which is given only by statistics
in an ideal detector, but in reality shows systematic effects which
mimic a genuine signal, and so have to be subtracted.

Systematic effects cannot be subtracted from the polarization degree and angle (which are not additive). This is one of the reasons for which it's more convenient to represent polarization through Stokes parameters; in X-ray polarimetry these are defined for each single event detected as \citep{2015APh....68...45K}
\begin{equation}
    q_i = 2\cos{(2\phi_i)}
\end{equation}
\begin{equation}
    u_i = 2\sin{(2\phi_i)}
\end{equation}
Spurious modulation is subtracted \citep{2022AJ....163...39R} from each single event based on its energy and its spatial position (found from the reconstructed absorption points of the events).

\begin{figure}
\begin{centering}
\includegraphics[width=1\linewidth]{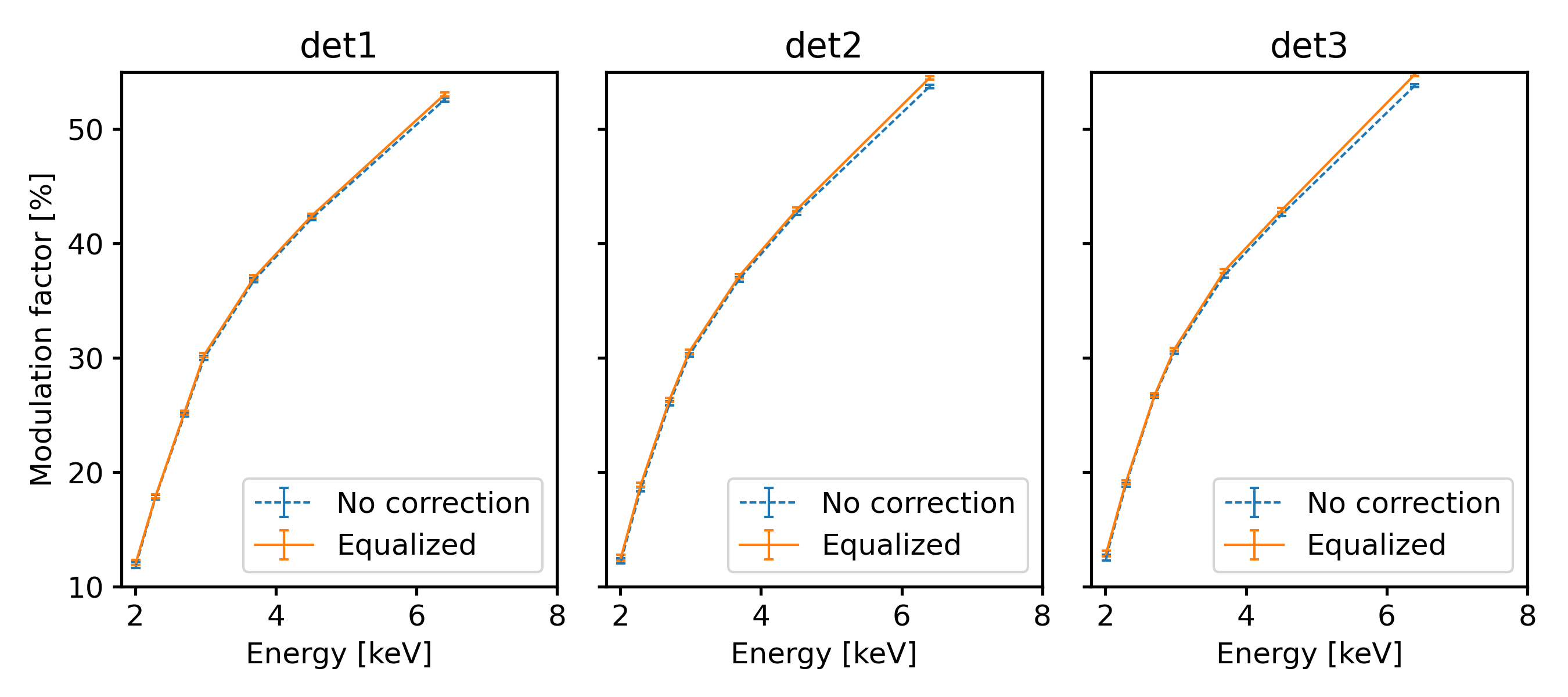}
\par\end{centering}
\caption{Modulation factor as a function of energy for the three detectors flying on-board IXPE. In all cases a slight improvement is visible at the highest energy when applying the pixel equalization. \label{fig:response_pol_en1}}
\end{figure}

\begin{figure}
\begin{centering}
\includegraphics[width=1\linewidth]{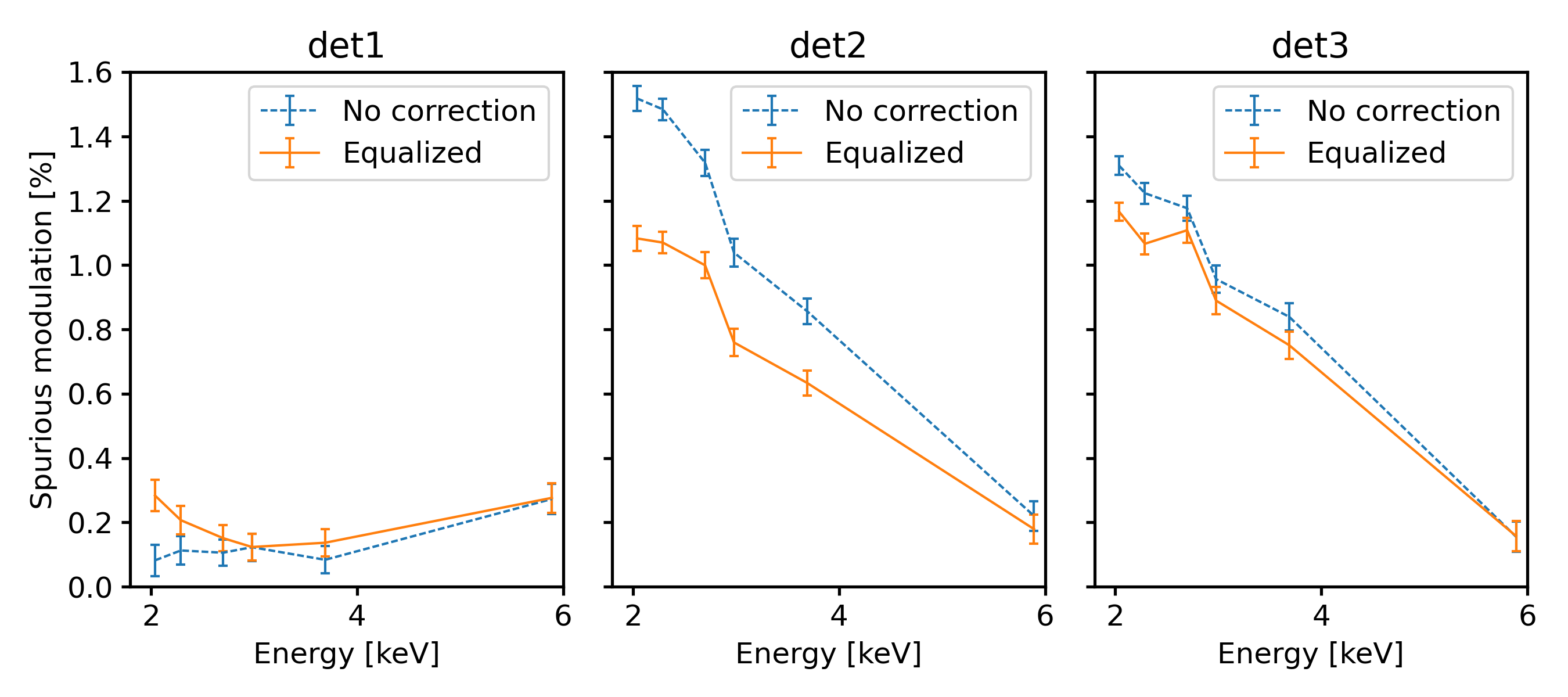}
\par\end{centering}
\caption{Spurious modulation as a function of energy for the three detectors flying on-board IXPE. Some improvement is
present when applying pixel equalization at low energy in detector 2 and 3; no improvement is visible for detector 1 for which spurious modulation is already very low. \label{fig:response_unpol_en1}}
\end{figure}

\begin{figure}
\begin{centering}
\includegraphics[width=1\linewidth]{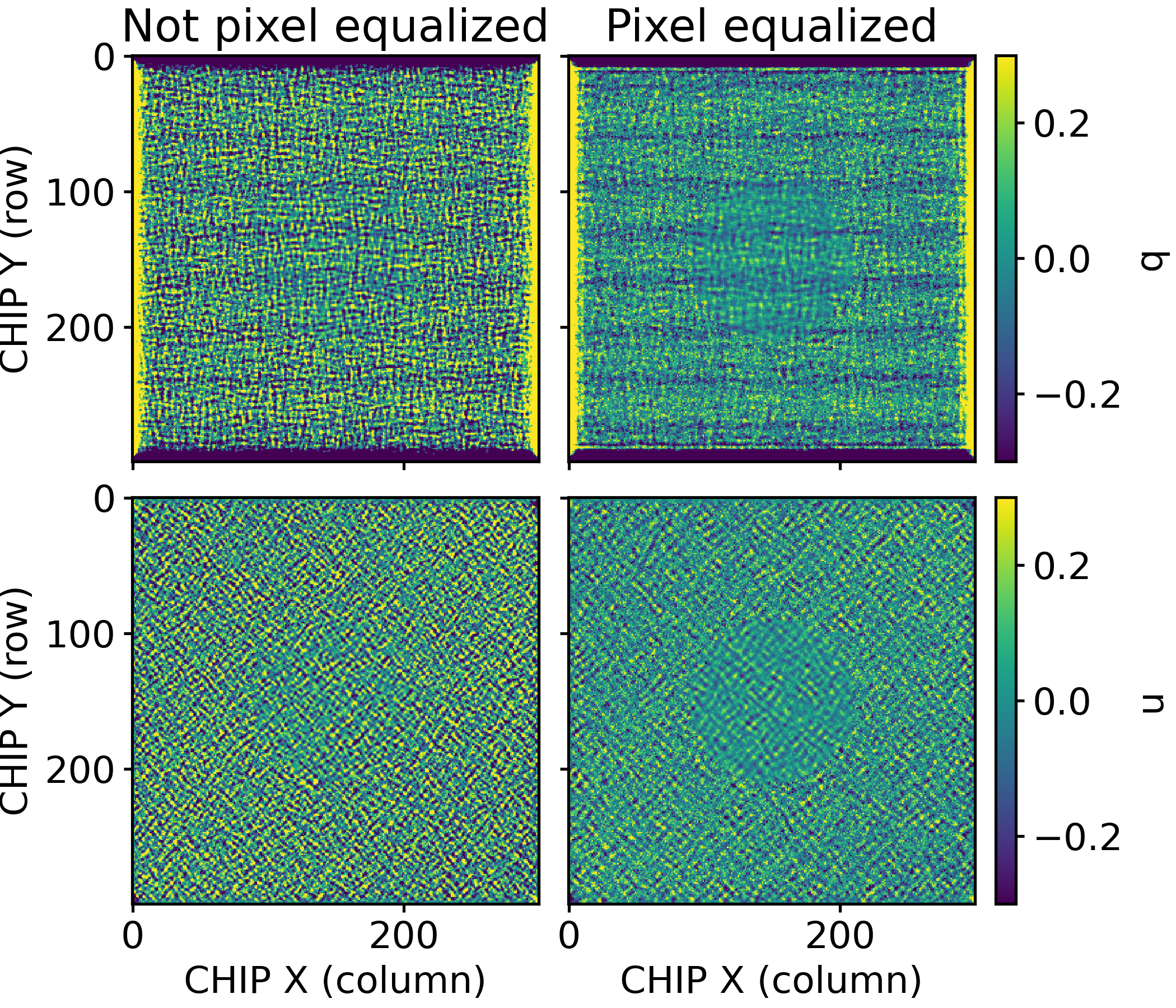}
\par\end{centering}
\caption{Spurious modulation maps at 2.7~keV generated with and without pixel
equalized data. The disuniformities are reduced when using equalized
data. \label{fig:spmod}}
\end{figure}

\begin{figure}
\begin{centering}
\includegraphics[width=1\linewidth]{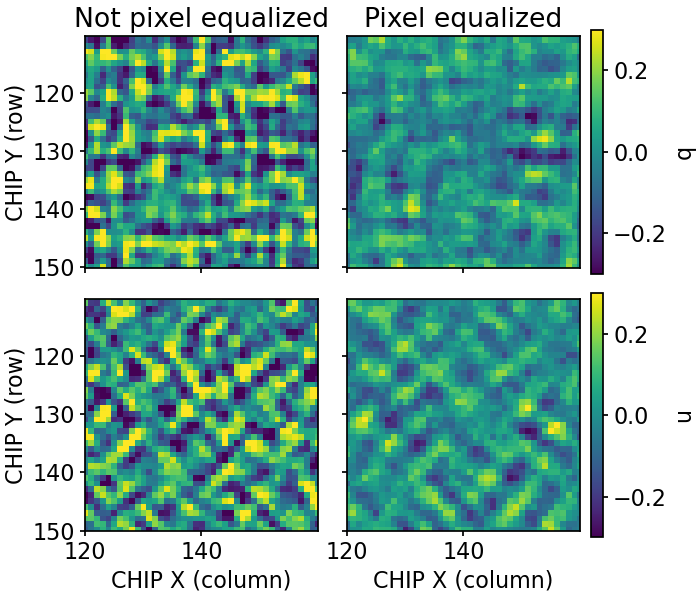}
\par\end{centering}
\caption{Same as figure \ref{fig:spmod} but zoomed-in to see the equalization
of the disuniformities. \label{fig:spmod-zoomed}}
\end{figure}

\begin{figure}
\begin{centering}
\includegraphics[width=1\linewidth]{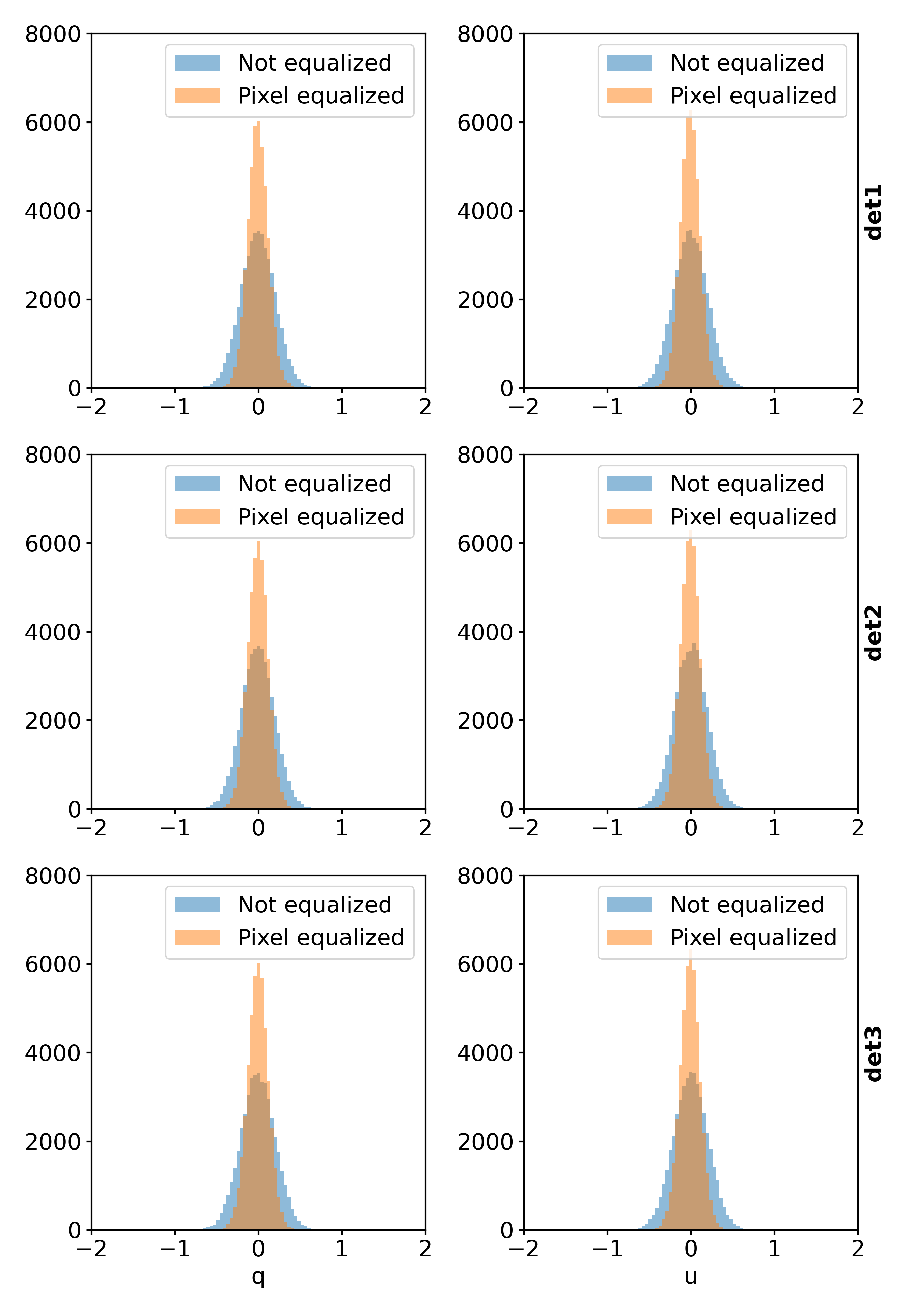}
\par\end{centering}
\caption{Same as figure \ref{fig:spmod} but as histogram, and for all three detectors flying on-board IXPE. The results for the three detectors are comparable. \label{fig:spmod-hist}}
\end{figure}

Figure \ref{fig:response_pol_en1} shows the modulation factor without and with pixel equalization, 
while figure \ref{fig:response_unpol_en1} shows spurious modulation, measured on a relatively-large spot of 1.5~mm radius.
The modulation factor shows a slight improvement, when applying the
pixel equalization, only at the highest energies.
For spurious modulation, a good improvement is visible
at low energy for detectors 2 and 3; no improvement is visible at high energy and for detector 1 because spurious modulation is already very low in these cases. The improvement
is up to 25\% at the lowest energies. But the greatest changes in spurious modulation, after
equalizing, are local --- on spatial scales comparable to the physical size of ASIC pixels, which is 50~$\mu$m. This can be seen in figures \ref{fig:spmod}, \ref{fig:spmod-zoomed}
and \ref{fig:spmod-hist}, where we show the map and the histograms of the Stokes parameters of spurious modulation in bins of 50$\times$50~$\mu$m$^2$. The amplitude of the effect is significantly reduced after equalization. This is expected: single pixels with larger or smaller gain can systematically shift the track charge distribution, and then the reconstructed direction of emission and the measured polarization. This effect tends to partially cancel out on larger spatial scales because the gain of different pixels essentially varies randomly. 

\section{Conclusion\label{sec:Conclusion}}

We presented a method to equalize the gain of the individual pixels of the ASIC specifically developed for the Gas Pixel Detector. Our approach relies on a peculiar functionality of this ASIC, which is able to identify and read-out only the relatively small region that records the signal, called ROI, instead of the entire matrix of pixels. The average charge collected by a specific pixel is compared with that collected on average by the others when they are in the same position within the ROI. 

This method was used to equalize the response of each pixel of the GPDs on-board IXPE, taking advantage of the calibration database built for the calibration of the polarization response of these instruments. Measurements at different energies, and results obtained with ROIs of different sizes, allow to sample the response of each specific pixel on an extensive range of the input charge. The response was measured to be essentially linear, with a slope changing up to $\pm$20\% for adjacent pixels. The offset is negligible, as expected from the fact that pedestals are subtracted in real-time during data acquisition with the GPD.

The equalization of pixel gain is beneficial for a number of detector performances. Energy resolution is greatly improved, as the uniformity of the energy response. Polarization response is mainly affected in the reduction of spurious modulation generated by detector disuniformities, especially at spatial scales comparable with the physical size of the pixel. Maps to equalize the pixel gain are currently being applied to all observations performed by the GPDs on-board IXPE. Nonetheless, this method could be applied in the future also to other missions with the GPD on-board, or to similar detectors able to resolve photoionization tracks in a sensitive medium.

\begin{acknowledgments}
The Italian contribution to the IXPE mission is supported by the Italian Space Agency (ASI) through the contract ASI-OHBI-2017-12-I.0, the agreements ASI-INAF-2017-12-H0 and ASI-INFN-2017.13-H0, and its Space Science Data Center (SSDC), and by the Istituto Nazionale di Astrofisica (INAF) and the Istituto Nazionale di Fisica Nucleare (INFN) in Italy. 

The United States contribution to the IXPE mission is supported by NASA as 
part of the Small Explorers Program. 
\end{acknowledgments}

\bibliography{equalization_citations}{}
\bibliographystyle{aasjournal}

\end{document}